\documentclass[10pt,conference]{IEEEtran}

\usepackage[utf8]{inputenc}
\usepackage{xspace}
\usepackage{xcolor}
\usepackage{listings}
\usepackage{balance}

\colorlet{punct}{red!60!black}
\definecolor{background}{HTML}{EFEFEF}
\definecolor{delim}{RGB}{20,105,176}
\colorlet{numb}{magenta!60!black}

\lstdefinelanguage{json}{
    basicstyle=\tiny\normalfont\ttfamily,
    numbers=none,
    numberstyle=\scriptsize,
    stepnumber=1,
    numbersep=8pt,
    showstringspaces=false,
    breaklines=true,
    frame=none,
    backgroundcolor=\color{background},
    literate=
     *{:}{{{\color{punct}{:}}}}{1}
      {,}{{{\color{punct}{,}}}}{1}
      {\{}{{{\color{delim}{\{}}}}{1}
      {\}}{{{\color{delim}{\}}}}}{1}
      {[}{{{\color{delim}{[}}}}{1}
      {]}{{{\color{delim}{]}}}}{1},
}

\usepackage[hidelinks, bookmarks=false]{hyperref}

\usepackage{tabularx}
\usepackage{booktabs}
\usepackage{multirow}
\usepackage{fp}
\usepackage[autolanguage]{numprint}
\newcommand*\np[2][z]{
\ifx z#1%
$\numprint{#2}$%
\else%
$\numprint[#1]{#2}$%
\fi\xspace
}

\usepackage{graphicx,pifont}
\let\oldding\ding
\renewcommand{\ding}[2][1]{\scalebox{#1}{\oldding{#2}}}
\usepackage{scalerel}
\usepackage{tikz}
\usetikzlibrary{svg.path}

\definecolor{orcidlogocol}{HTML}{A6CE39}
\tikzset{
	orcidlogo/.pic={
		\fill[orcidlogocol] svg{M256,128c0,70.7-57.3,128-128,128C57.3,256,0,198.7,0,128C0,57.3,57.3,0,128,0C198.7,0,256,57.3,256,128z};
		\fill[white] svg{M86.3,186.2H70.9V79.1h15.4v48.4V186.2z}
		svg{M108.9,79.1h41.6c39.6,0,57,28.3,57,53.6c0,27.5-21.5,53.6-56.8,53.6h-41.8V79.1z M124.3,172.4h24.5c34.9,0,42.9-26.5,42.9-39.7c0-21.5-13.7-39.7-43.7-39.7h-23.7V172.4z}
		svg{M88.7,56.8c0,5.5-4.5,10.1-10.1,10.1c-5.6,0-10.1-4.6-10.1-10.1c0-5.6,4.5-10.1,10.1-10.1C84.2,46.7,88.7,51.3,88.7,56.8z};
	}
}

\newcommand\orcidicon[1]{\href{https://orcid.org/#1}{\mbox{\scalerel*{
				\begin{tikzpicture}[yscale=-1,transform shape]
				\pic{orcidlogo};
				\end{tikzpicture}
			}{|}}}}
			
\AtBeginDocument{%
}

\newcommand{\ShowAbsoluteNumber}[1]{%
\ifnum #1<10%
{\hspace*{0pt}#1}%
\else%
\ifnum #1<100%
{\hspace*{0pt}#1}%
\else%
\ifnum #1<1000%
{\hspace*{0pt}#1}%
\else%
{\numprint{#1}}%
\fi%
\fi%
\fi%
}

\newcommand{\ShowPercentage}[2]{%
\FPeval\percentage{round(#1/#2*10000,0)}%
\FPeval\percentageOneDecimal{round(#1/#2*100,1)}%
\ifnum \percentage=0%
    {\np[\%]{0}}%
\else%
    \ifnum \percentage<10%
        {$<$\np[\%]{0.1}}%
    \else%
        {\np[\%]{\FPprint{percentageOneDecimal}}}%
    \fi%
\fi%
\xspace
}
\newlength\BARSIZE  
\newcommand{\inlinechart}[2]{%
\FPeval{\BLACKBARSIZE}{#1/#2}\textcolor{black!80}{\rule{\BLACKBARSIZE\BARSIZE}{1.6ex}}%
\FPeval{\BLACKBARSIZE}{1 - (#1/#2)}\textcolor{black!10}{\rule{\BLACKBARSIZE\BARSIZE}{1.6ex}}%
}

\newcommand*\ChartSmall[3][v]{%
\ifx q#1%
    \np{#2}/\np{#3}(\ShowPercentage{#2}{#3})\else%
\ifx p#1%
    \np{#2}(\ShowPercentage{#2}{#3})\else%
\ifx c#1%
    \inlinechart{#2}{#3}%
\else%
    \np{#2}%
    \ifx r#1%
        /\np{#3}%
    \fi%
    \hspace*{0.5ex}(\ShowPercentage{#2}{#3}) %
    \inlinechart{#2}{#3}%
    \xspace
\fi\fi\fi%
}
\def\nbLib{94}
\def\nbLibStr{\np{\nbLib}}
\def\nbLibVersion{395}
\def\nbLibVersionStr{\np{\nbLibVersion}}

\def\nbTest{713932}
\def\nbTestStr{\np{\nbTest}}
\def\nbTestClient{211116}
\def\nbTestClientStr{\np{\nbTestClient}}
\def\nbClientAll{2874}
\def\nbClientAllStr{\np{\nbClientAll}}

\def\medianCoverageLib{80.83}
\def\medianCoverageLibStr{\np[\%]{\medianCoverageLib}}

\def\medianCoverageClient{20.24}
\def\medianCoverageClientStr{\np[\%]{\medianCoverageClient}}
\def\nbLineLib{10831394}
\def\nbLineLibStr{\np{\nbLineLib}}
\def\nbLineClient{140910102}
\def\nbLineClientStr{\np{\nbLineClient}}

\def\nbLibDebSuccessNum{311}

\FPeval{nbDebloatError}{round(\nbLibVersion - \nbLibDebSuccessNum, 0)}

\def\nbLibDebSuccessNum{311}
\def\nbLibPassTestNum{220}

\def\nbLibTestFailing{30}
\def\nbUniqueTestNum{344596}

\def\nbFailingTest{1405}
\def\nbErrorTest{432}

\FPeval{nbLibExecTestNum}{round(\nbLibPassTestNum + \nbLibTestFailing, 0)}
\FPeval{nbPassingTestNum}{round(\nbUniqueTestNum - \nbFailingTest, 0)}
\FPeval{nbFailureTest}{round(\nbFailingTest - \nbErrorTest, 0)}

\def\nbLibWithDependencies{77}
\def\nbDependencies{254}

\def\nbBloatedDependencies{52}

\FPeval{nbLibWithoutDependencies}{round(\nbLibPassTestNum - \nbLibWithDependencies, 0)}
\FPeval{nbNonBloatedDependencies}{round(\nbDependencies - \nbBloatedDependencies, 0)}

\def\bytecodeSize{893732414}
\def\debloatedSize{596538866}
\def\debloatedMethodSize{13768627}

\FPeval{bloatedSize}{round(\debloatedSize + \debloatedMethodSize, 0)}
\FPeval{nonBloatedSize}{round(\bytecodeSize - \debloatedSize - \debloatedMethodSize, 0)}

\def\nbDynCoveringClient{283}
\def\nbStatCoveringClient{1001}

\def\nbFailingTestClient{54}

\def\nbClientDebloatError{44}

\def\nbTotalProject{147991}
\def\nbTotalProjectStr{\np{\nbTotalProject}}

\FPeval{nbCientDebloatNoCompError}{round(\nbStatCoveringClient - \nbClientDebloatError, 0)}
\FPeval{nbClientDebloatNoTestFail}{round(\nbDynCoveringClient - \nbFailingTestClient, 0)}

\FPeval{nbFailingTestNum}{round(\nbUniqueTestNum - \nbPassingTestNum, 0)}
\FPeval{nbSuccessTestNum}{round(\nbPassingTestNum*100/\nbUniqueTestNum, 2)}

\FPeval{nbLibFailTestNum}{round(\nbLibDebSuccessNum - \nbLibPassTestNum, 0)}

\newcommand{\ie}{i.e.\@\xspace}

\newcommand{\eg}{e.g.\@\xspace}

\newcommand{\etal}{et al.\@\xspace}

\newcommand\pom{\texttt{pom.xml}\xspace}

\newcommand\duets{{\textsc{Duets}}\xspace}

\newcommand\mv{{Maven}\xspace}

\title{\duets: A Dataset of Reproducible Pairs of\\ Java Library-Clients}
\author{
\IEEEauthorblockN{
	\normalsize
	Thomas Durieux$^{~\orcidicon{0000-0002-1996-6134}}$,
	C\'esar Soto-Valero$^{~\orcidicon{0000-0003-0541-6411}}$,
	   and Benoit Baudry$^{~\orcidicon{0000-0002-4015-4640}}$}

\IEEEauthorblockA{
\textit{KTH Royal Institute of Technology, Stockholm, Sweden}\\
Email: \{tdurieux, cesarsv, baudry\}@kth.se\\
}
}

\IEEEoverridecommandlockouts

\begin{document}

\maketitle

\begin{abstract}
Software engineering researchers look for software artifacts to study their characteristics or to evaluate new techniques. In this paper, we introduce \duets, a new dataset of software libraries and their clients. 
This dataset can be exploited to gain many different insights, such as API usage, usage inputs, or novel observations about the test suites of clients and libraries. 
\duets is meant to support both static and  dynamic analysis. This means that the libraries and the clients compile correctly, they are executable and their test suites  pass.
The dataset is composed of open-source projects that have more than five stars on GitHub. The final dataset contains \nbLibVersionStr libraries and \nbClientAllStr clients. 
Additionally, we provide the raw data that we use to create this dataset, such as \np{34560} \pom files or the complete file list from \np{34560} projects.
This dataset can be used to study how libraries are used by their clients or as a list of software projects that successfully build. The client's test suite can be used as an additional verification step for code transformation techniques that modify the libraries.
\end{abstract}


\section{Introduction}

\IEEEPARstart{S}{oftware} engineering research requires real software artifacts either to study their properties or to evaluate new techniques. Software datasets have emerged in the community as an effort to standardize and increase the reproducibility of software studies and the comparison between contributions. 
Each dataset focuses on one specific goal and has specific properties.
For example, some datasets focus on the source code of different projects \cite{Pietri2019}, others focus on software that  compiles correctly \cite{Martins2018}, or even focus on specific characteristics such as having known and reproducible bugs \cite{Just2014, Madeiral2019}.

This paper presents a new dataset of software projects: \duets.
Its name reflects the spirit of the  library-client relationship.
It consists of a collection of Java libraries, which build can be successfully reproduced, and Java clients that use those libraries.
\duets aims to simplify research that focuses on behavioral analysis of Java software. In particular we want to encourage studies that  analyze library behavior in a context. 
The availability of a set of clients for each library supports studies about the actual usage of the library. 
The dataset can be used both for static analyses and for dynamic analyses by executing the tests of the clients. 
For that purpose, we take a special care to build a dataset for which we ensure that both the library and the clients have a passing test suite.

This new dataset supports a wide range of usage purposes. In general, many program analyses require a list of compilable and testable software packages, which we provide with \duets. The dataset also supports  more specific use cases, such as analyzing the API usage by the clients of a particular library \cite{Saied2015}, or debloating based on dynamic analysis \cite{jdbl, JShrink}.
We provide a framework with the dataset. It can be used, for example, to detect projects that have flaky builds or study the reasons for flakiness or the build~\cite{Pinto2020}.
In addition, the reproducible property of the project builds provides a sound ground for empirical studies on how \pom files (Maven build configuration file) are engineered \cite{Wang2018,McIntoshAH12, Valero2020}.

We design \duets to contain a large diversity of projects and focus on the reproducibility of the build.
We select only single-module Maven projects to simplify the reproducibility. 
Multi-module projects tend to increase the complexity and the fragility of the build and make it harder to analyze, study, and instrument. 
We also build each project three times, as an effort to ensure the reproducibility of the build.
The test suite of the project has to pass and only have passing tests.

To summarize the contributions of this paper are the follow:
\begin{itemize}
    \item \duets, a dataset of \nbLibVersionStr libraries and \nbClientAllStr clients. Both the libraries and the clients build successfully with Maven, \ie all the test pass and a compiled artifact is produced as a result of the build.
    \item A framework to generate the dataset. It includes scripts for the collection of data and execution of the tests in a sandboxed environment.
    \item Raw data files from the dataset generation that can be reused for further researches, e.g., \np{34560} \pom files, \np{33513} Travis CI configuration files,  and \np{17403} Dockerfiles. 
\end{itemize}

\section{Data Collection Methodology}

\begin{figure*}[t]
	\centering
	\includegraphics[origin=c,width=0.90\textwidth]{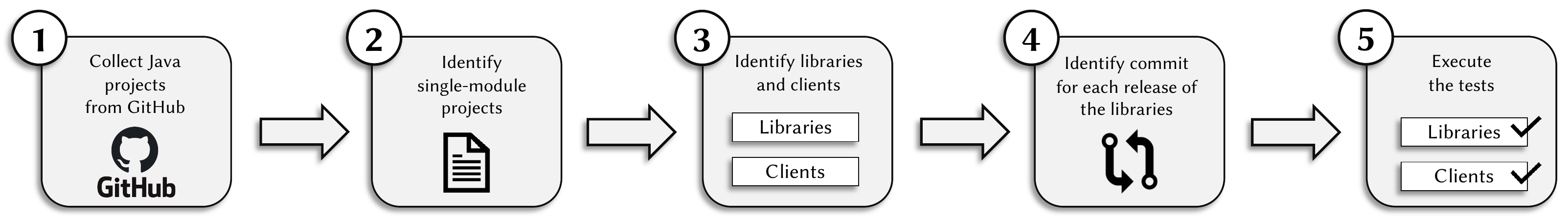}
	\caption{Overview of the data collection methodology: from a collection of Java projects on GitHub, it produces reproducible sets of libraries and clients.}
	\label{fig:overview}
	\vspace{-0.25cm}
\end{figure*}

In this section, we describe the methodology that we follow to construct this dataset of open-source \mv Java projects extracted from GitHub.
\duets is composed of two parts: a set of libraries, \ie, Java projects that are declared as a dependency by other Java projects, and a set of clients, \ie, Java projects that use the libraries from the first set.
The construction of this dataset is performed in $5$ steps. 
The process is illustrated in \autoref{fig:overview} and detailed in the following sections. 

\subsubsection{Collection of Java projects from GitHub}

First, we use the list of Java projects extracted from GitHub by Loriot et al. \cite{styler}. The authors queried the GitHub API on June 9th of 2020, to find all the projects that use Java as the primary programming language. 
The projects found were subsequently filtered, discarding those that have less than $5$ stars.
This initial dataset includes  \nbTotalProjectStr Java projects.

\subsubsection{Identification of single-module projects}

Second, we select the subset of single-module projects among the \nbTotalProjectStr~Java projects.
We choose single-module projects to have a clear mapping between client and library and have  more reliable build reproduction.

We download the complete list of files for each project using the GitHub API.
We consider that a project is a single-module project when it has  a single \mv build configuration file, \ie, \pom.
We exclude the \pom files that are in \texttt{resource} folders and \texttt{test} folders.
At the end of this step, we keep \ChartSmall[p]{34560}{\nbTotalProject} single-module \mv projects.
The list of all the files is also part of our dataset and is available in the repository of the dataset.

\subsubsection{Identification of libraries and clients}

In the third step, we analyze each \pom from the \np{34560} projects.
During the analysis, we first extract the \texttt{groupId}  and \texttt{artifactId} qualifiers of each project. 
This pair of ids is used by \mv to identify a project.
In the case where two projects declare the same pair of \texttt{groupId} and \texttt{artifactId}, we select the project that has the largest number of stars on GitHub.
Second, we map the \texttt{groupId} and \texttt{artifactId} to the dependency declared in the \pom.
At the end of this step, we obtain a list of projects that are used as dependency, \ie, libraries and a list of clients that use the libraries.
During this step, we ignore the projects that do not declare JUnit as a testing framework, and we exclude the projects that do not declare a fixed release, \eg, \texttt{LAST-RELEASE}, \texttt{SNAPSHOT}.
After this third step, we identify \ChartSmall[p]{155}{34560} libraries, and \ChartSmall[p]{25557}{34560} clients that use \np{2103} versions of the libraries.

\subsubsection{Identification of commits for each library release}

The purpose of the fourth step is to identify the commit SHA identifier that determines each version of the library, \ie, the commit change in the \pom that assigns a new version.
For example, the version $3.4$ of the library \texttt{commons-net} is defined in the commit SHA \texttt{74a2282b7e4c6905581f4f1b5a2ec412310cd5e7}.
To perform this task, we download all revisions of the \mv build configuration files since their creation. 
Then, we analyze the \mv build configuration files, and identify which commit declares  a specific release of the library.
We successfully identify the commit for \ChartSmall[q]{1026}{2103} versions for \ChartSmall[q]{143}{155} libraries. \ChartSmall[q]{16964}{25557} clients have been mapped to a specific commit of one of their dependencies.
\looseness=-1

\subsubsection{Execution of the tests}\label{sec:reproduice}

As the fifth and last step, we execute three times the test suite of all library versions and all clients, as a sanity check to filter out libraries with flaky tests or projects that cannot be built.
We keep the libraries and clients that have at least one test and have all tests passing: \ChartSmall[q]{94}{143} libraries, \ChartSmall[q]{395}{1026} library versions, and \ChartSmall[q]{2874}{16964} clients passed this verification.
From this point, we consider each library version as a unique library for clarification purpose.

\section{Description of the dataset}

\begin{table*}[t]
\centering
    \caption{Descriptive statistics of the studied libraries and their associated clients}
    \label{tab:benchmark}
    \begin{tabularx}{.95\textwidth}{@{}l X rrrrrrr@{}}
    \toprule
    &                        & Min & 1st Qu. & Mean & 3rd Qu. & Max & Avg. & Total \\
    \midrule
\multirow{7}{*}{\nbLibVersionStr Libraries}
& \# Line of code   & \np{132} & \np{5439.5} & \np{17935.5} & \np{47866.0} & \np{341429} & \np{35629.6} & \nbLineLibStr \\
& \# Contributors & \np{1} & \np{14.25} & \np{29} & \np{58.5} & \np{286} & \np{48.79} & \np{4586} \\
& \# Years of activity & \np{0.35} & \np{6.25} & \np{9.49} & \np{15.21} & \np{22.35} & \np{10.32} & N.A \\
& \# Commits & \np{2} & \np{217.25} & \np{623} & \np{1538.75} & \np{16570} & \np{1283.46} & \np{120645} \\
& \# Stars & \np{5} & \np{72.75} & \np{266.5} & \np{701.5} & \np{21865} & \np{1457.82} & \np{137035} \\
& \# Tests & \np{1} & \np{139.8} & \np{378.0} & \np{1108.2} & \np{24946} & \np{1830.6} & \nbTestStr \\
& Coverage & \np[\%]{0.1} & \np[\%]{61.7} & \np[\%]{80.8} & \np[\%]{89.8} & \np[\%]{100.0} & \np[\%]{73.7} & N.A \\
\hline
\multirow{7}{*}{\nbClientAllStr Clients} 
& \# Line of code   & \np{0} & \np{3130.0} & \np{9170.0} & \np{58990.0} & \np{4531710} & \np{72897.1} & \nbLineClientStr \\
& \# Contributors & \np{1} & \np{2} & \np{4} & \np{11} & \np{286} & \np{10.98} & \np{31493} \\
& \# Years of activity & \np{0} & \np{0.94} & \np{2.65} & \np{4.92} & \np{23.80} & \np{3.35} & N.A \\
& \# Commits & \np{1} & \np{27} & \np{86} & \np{261} & \np{18594} & \np{370.26} & \np{1062286} \\
& \# Stars & \np{5} & \np{7} & \np{14} & \np{40} & \np{21865} & \np{118.96} & \np{341287} \\
& \# Tests & \np{1} & \np{4.5} & \np{20.0} & \np{74.0} & \np{11415} & \np{107.7} & \nbTestClientStr \\
& Coverage & $<$\np[\%]{0.1} & \np[\%]{2.1} & \np[\%]{20.24} & \np[\%]{57.7} & \np[\%]{100.0} & \np[\%]{31.4} & N.A \\
\bottomrule
\end{tabularx}
\end{table*}
 
\autoref{tab:benchmark} summarizes the descriptive statistics of the dataset.
The number of lines of code (\#LOC) and the coverage are computed with JaCoCo.
\duets includes \nbLibStr different libraries, with a total of \nbLibVersionStr versions, as well as  \nbClientAllStr clients. 
Those libraries and clients are maintained by \np{1669} different GitHub organizations.
The libraries include \nbTestStr test cases that cover \medianCoverageLibStr of the \nbLineLibStr LOC. The libraries have a median  maintenance time of $9.49$ years from $623$ commits created by $29$ contributors.
The clients have \nbTestClientStr test cases that cover \medianCoverageClientStr of the \nbLineClientStr LOC. The clients have a median  maintenance time of $2.65$ years from $86$ commits created by $4$ contributors.
The dataset and the scripts to generate the dataset are publicly available in our experiment repository: \url{https://github.com/castor-software/Duets}.

\subsection{Dataset format}

\duets is available on GitHub and is composed of a JSON file.\footnote{\url{https://github.com/castor-software/Duets/tree/master/dataset/dataset-info.json}}
An excerpt of this JSON file is presented in the README of our repository.
It contains the repository name, the SHA of the commit, the \texttt{groupId}, \texttt{artifactId}, the list of clients for each version of the library, and a list of commits that defines the different releases of the library.
\duets also contains the logs corresponding to the test execution for each version of the libraries and for each client. 
As previously mentioned, we executed three times the tests to increase the likelihood of the reproducibility of the dataset.

In addition to the dataset itself, we include all scripts that generate the dataset. Those scripts can be used to reproduce the same dataset, to create a similar dataset for a different language or reusing too mine Github.

\subsection{Execution framework}
In addition to the JSON file, we provide a Docker image that contains our execution framework.
This framework adds an abstraction on top of Git repositories. It automatizes the cloning, checkout, execution of the test, and parsing the test results without requiring to specify and additional information than the URL of the repository.
All those tasks are simplified into the following command line: \texttt{docker run --rm -v `pwd`:/results castorsoftware/duets:latest compile --repository https://github.com/radsz/jacop --commit 8f09fd977a}.
This framework can be extended to perform additional tasks, we provide an API that allows to execute the main Maven tasks and to manipulate the \pom files easily (\eg, to add or remove plugins). We extended the framework to perform static analysis and to collect the test suite coverage. Those examples are provided in the \duets repository as guidelines.

\section{Dataset Usage}

\subsection{Pairs of libraries and clients}

\begin{figure}[t]
	\centering
	\includegraphics[origin=c,width=0.40\textwidth]{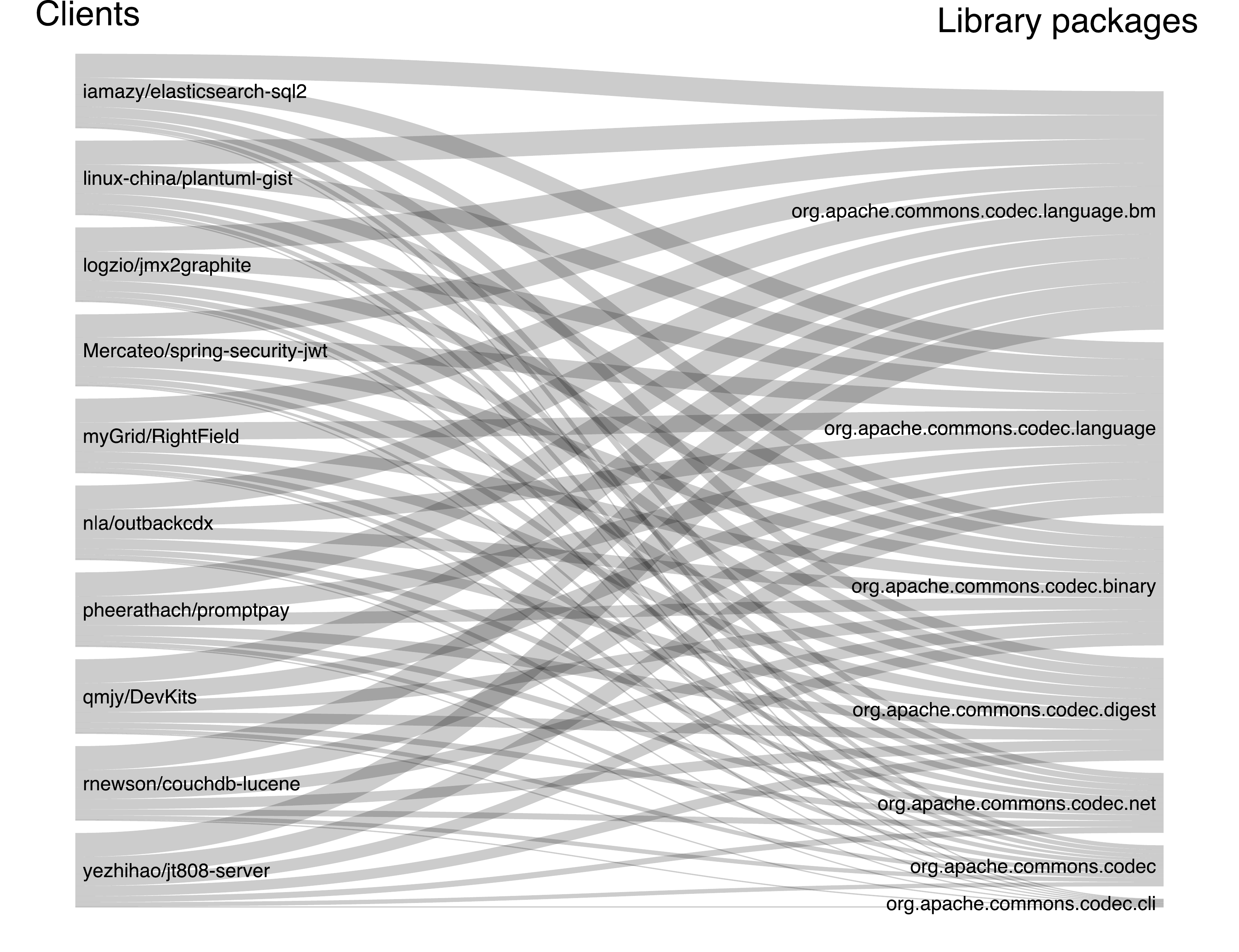}
	\caption{Sankey diagram representing the number of classes in the packages of the library Apache Commons Codec (v. $1.12$) that are used by $10$ clients}
	\label{fig:sankey}
	\vspace{-0.3cm}
\end{figure}

The clients in \duets can be used to identify APIs usage patterns between different clients \cite{Saied2015}, or to explore how the API evolution of the libraries affects their clients \cite{Eilertsen2018}. 
For example, \autoref{fig:sankey} shows a weighted bipartite graph of the relationship between $10$ clients (on the left) and the packages belonging to one  library, Apache Commons Codec (on the right).
The width of the edges between a client and a package  represents the number of classes in the package that are executed (either directly called or invoked internally in the library) when running the clients' tests. 
This type of figure allows to visualize the parts of the library that are more used by its clients, which is valuable information for both the library users and the library maintainers.

The test-suite of the clients can also be used as further validation of modifications  performed on the library. 
This is the type of usage that we leverage in our recent work \cite{jdbl}, where we debloat libraries and verify that the compilation and the execution of their clients' tests are not affected.
Hence, having information regarding library usage by clients is useful for validating program transformation since it provides dynamic data that helps to overcome the limitations of static analysis in Java.
\looseness=-1

Another potential usage of the clients of the library is to use the client tests to generate tests for the libraries or to verify that changes in the libraries do not break the clients. This idea is developed in a recent study performed by Chen and colleagues \cite{Chen2020}.

\duets allows to compare the characteristics of libraries with respect to  other software artifacts. For example, during our data collection, we observed that libraries have much more tests and have a higher test coverage than the clients (see \autoref{tab:benchmark}). 
A detailed analysis of the test part of the dataset could highlight differences between libraries and other types of applications.

Finally, \duets can be used to compare the coverage of the library with its own test suite and the test suites of the clients, to identify the intersection and difference between the two test suites, similar to the work of Wang \etal~\cite{wang2017behavioral} that checks the similarity between test and production behavior.

\subsection{Buildable and testable Java projects}

If the relations between the clients and the libraries are not required for a specific evaluation. 
\duets can be used as a list of projects that successfully build and have a passing test suite.
This can be used as a dataset for dynamic analysis such as identifying API usage based on the client or libraries test suites.
This dataset contains a large diversity of projects, large and small, from different fields.

\subsection{Build results of the projects}

During the creation of \duets, we verified that \np{7293} projects are reproducible and have only passing tests (see \autoref{sec:reproduice}). We saved the test results of those executions.
This data can be used to identify projects with failing tests, flaky tests or flaky builds.
Identifying flaky builds is a hard task. This data could have simplified the work of studies like  \cite{luo2014empirical,eck2019understanding,durieux2020flaky}.
\autoref{tab:builds_metrics} presents some metrics of our reproduction attempts. 
We identify \ChartSmall[p]{221}{7293} projects that have flaky tests, \ChartSmall[p]{1009}{7293} projects with at least one failing test case.
\vspace{-1.5em}
\begin{table}[h!]
    \centering
    \caption{Reproducibility metrics report}
    \label{tab:builds_metrics}
    \begin{tabular}{@{}lr@{}}
    \toprule
Metrics    &  Value\\\midrule
\# Reproduction attempts & \np{7293}\\
\# Buidable projects & \ChartSmall{3642}{7293}\\
\# Unbuidable projects & \ChartSmall{3418}{7293}\\
\# Failing-test builds & \ChartSmall{1009}{7293}\\
\# Flaky builds & \ChartSmall{221}{7293}\\
\# Timeout & \ChartSmall{12}{7293}\\
    \bottomrule
    \end{tabular}
\end{table}

\vspace{-1.5em}
\subsection{List of files of Java projects}

We downloaded the complete list of files for \np{34560} Java projects from GitHub. This list of files can be used to identify the usage of specific technologies such as Docker, continuous integration, build management systems, or investigating the adoption of  some development practice such as including binaries in the repositories. 
\autoref{tab:project-files} shows the rate of occurrence of these particular files in \duets.
This data could be used for study like the one of Cito \etal \cite{cito2017empirical}.
\begin{table}[h!]
\vspace{-0.25cm}
    \caption{Occurrence of Java files and configuration files in the dataset of file lists}
    \label{tab:project-files}
    \centering
    \begin{tabular}{@{}lr@{}}
    \toprule
Metrics    &  Values\\\midrule
\# Files & \np{71768708} \\
\# Java files & \ChartSmall{21519119}{71768708} \\
\# \pom files & \ChartSmall{363220}{71768708} \\
\# Gradle files & \ChartSmall{229690}{71768708} \\
\# Travis files & \ChartSmall{33513}{71768708} \\
\# GitHub Workflow files & \ChartSmall{33513}{71768708} \\
\# Dockerfiles & \ChartSmall{17403}{71768708} \\
\bottomrule
\end{tabular}
\end{table}
\vspace{-1.5em}
\subsection{Analysis of \pom files}

\duets contains \np{34560} \pom files. Those files can be used to analyze the common usage of \pom in open source repositories.
This dataset of \pom files has several advantages compared to a dataset of \pom created directly from Maven Central.
The \pom in \duets are directly associated with a Git repository and therefore additional information is available, such as the source code, the history of the project, or issues. This provides a solid starting point to analyze the co-evolution of build files and other artifacts \cite{mcintosh2014mining}.

\section{Related Work}

Building software is a complex task. 
Indeed, Kerzazi \etal \cite{kerzazi2014automated} observe that $17.9\%$ of CI builds in an industrial web application are failing during a period of $6$ months. 
Durieux \etal \cite{durieux2020flaky} shows that only $70\%$ of the builds are passing on Travis CI.
Reproducing builds is even an harder task.  
Sulir \etal \cite{sulir2016quantitative} show that $40\%$ of the builds in their dataset are not reproducible. Almost $40\%$ of the build problems are related to missing dependencies, followed by compilation errors in $22\%$ of the cases.  
Gkortzis \etal
\cite{gkortzis2020software} observe a very similar build failure rate ($33.7\%$), with the same causes. 
Neitsch \etal \cite{neitsch2012build} analyze the build systems of $5$ open-source multi-language Ubuntu packages. 
They observe that $4$ of the $5$ packages cannot build or be rebuilt. They find that many build problems can be addressed, and note that build quality  is rarely the subject of research, even though it is an important part of maintaining and reusing software.

The software engineering research community came up with several benchmarks of Java projects that focus on reproducibility of the build.
Sulir \etal \cite{sulir2016quantitative} attempted to build \np{7264} Java projects from GitHub, from which around $60\%$ of the builds succeeded.
Martins \etal~\cite{Martins2018} presented in 2018 a dataset that follows the same idea but with \np{50000} compilable and compiled Java projects. 
Dacapo by Blackburn \etal~\cite{Dacapo}  consists of a set of open source, real world applications with non-trivial memory loads.
The difference between those datasets and \duets is that we constructed a up-to-date benchmark with recent and diverse projects, we also focus on Mavven that have a test suite and all  tests are passing.

There are several datasets of buggy Java programs that generally also come with the non-buggy version of the program, such as \cite{Just2014,Madeiral2019,tomassi2019bugswarm,saha2018bugs,durieux2016introclassjava}.
Datasets that only focus on source code also exist, such as Boa, a dataset of queryable Java AST presented by Dyer \etal \cite{dyer2013boa}. Spinellis \etal \cite{spinellis2020dataset} focus on identifying duplicated repositories on GitHub.

The closest work that focuses on studying libraries and their clients is the work from Leuenberger \etal \cite{Leuenberger2017}.
They analyze the binaries of artifacts in Maven Central to identify API clients.
In contrast, we focus on the projects' source code and to allow to build and test the software where Leuenberger \etal are interested to mine the API usage of compiled projects.

The major difference between all those datasets and \duets is that we focus on pairs of libraries and clients. To our knowledge, this has never been done. 

\section{Conclusion}

In this paper, we presented \duets, a dataset of \nbLibVersionStr libraries and \nbClientAllStr clients extracted from open-source projects on GitHub.
\duets aims to simplify studies that rely on dynamic and static analysis of libraries' usage in the Java ecosystem.
In our previous work, we have used this dataset to study the impact of debloatig libraries on their clients.
However, \duets also provides a fertile ground for other types of empirical studies, such as those that analyze the impact of API changes on library clients.
Alongside the dataset itself, we provide a framework that aims to facilitate the data mining.
Both the dataset and the necessary tools to reproduce it are open-source and publicly available online.
We also provide the raw data that we use to generate \duets, including \np{34560} \pom files and the complete file list of \np{34560} Java projects.

\section*{Acknowledgment}

This work is partially supported by the Wallenberg AI, Autonomous Systems, and Software Program (WASP) funded by Knut and Alice Wallenberg Foundation and by
the TrustFull project funded by the Swedish Foundation for
Strategic Research.\looseness=-1

\balance
\bibliographystyle{ieeetr}
\bibliography{biblio}

\begin{thebibliography}{10}

\bibitem{Pietri2019}
A.~Pietri, D.~Spinellis, and S.~Zacchiroli, ``The software heritage graph
  dataset: Public software development under one roof,'' in {\em Proceedings of
  the 16th International Conference on Mining Software Repositories}, MSR '19,
  p.~138–142, IEEE Press, 2019.

\bibitem{Martins2018}
P.~Martins, R.~Achar, and C.~V. Lopes, ``50k-c: A dataset of compilable, and
  compiled, java projects,'' in {\em 2018 IEEE/ACM 15th International
  Conference on Mining Software Repositories (MSR)}, pp.~1--5, IEEE, 2018.

\bibitem{Just2014}
R.~Just, D.~Jalali, and M.~D. Ernst, ``{Defects4J}: A {Database} of existing
  faults to enable controlled testing studies for {Java} programs,'' in {\em
  ISSTA 2014, Proceedings of the 2014 International Symposium on Software
  Testing and Analysis}, (San Jose, CA, USA), pp.~437--440, July 2014.
\newblock Tool demo.

\bibitem{Madeiral2019}
F.~{Madeiral}, S.~{Urli}, M.~{Maia}, and M.~{Monperrus}, ``Bears: An extensible
  java bug benchmark for automatic program repair studies,'' in {\em 2019 IEEE
  26th International Conference on Software Analysis, Evolution and
  Reengineering (SANER)}, pp.~468--478, 2019.

\bibitem{Saied2015}
M.~A. Saied, O.~Benomar, H.~Abdeen, and H.~Sahraoui, ``Mining multi-level api
  usage patterns,'' in {\em 2015 IEEE 22nd international conference on software
  analysis, evolution, and reengineering (SANER)}, pp.~23--32, IEEE, 2015.

\bibitem{jdbl}
C.~{Soto-Valero}, T.~{Durieux}, N.~{Harrand}, and B.~{Baudry}, ``{Trace-based
  Debloat for Java Bytecode},'' {\em arXiv e-prints}, p.~arXiv:2008.08401, Aug.
  2020.

\bibitem{JShrink}
B.~R. Bruce, T.~Zhang, J.~Arora, G.~H. Xu, and M.~Kim, ``Jshrink: In-depth
  investigation into debloating modern java applications,'' in {\em Proceedings
  of the 28th ACM Joint Meeting on European Software Engineering Conference and
  Symposium on the Foundations of Software Engineering}, ESEC/FSE 2020, (New
  York, NY, USA), p.~135–146, Association for Computing Machinery, 2020.

\bibitem{Pinto2020}
G.~Pinto, B.~Miranda, S.~Dissanayake, M.~d'Amorim, C.~Treude, and A.~Bertolino,
  ``What is the vocabulary of flaky tests?,'' in {\em Proceedings of the 17th
  International Conference on Mining Software Repositories}, MSR '20, (New
  York, NY, USA), p.~492–502, Association for Computing Machinery, 2020.

\bibitem{Wang2018}
Y.~Wang, M.~Wen, Z.~Liu, R.~Wu, R.~Wang, B.~Yang, H.~Yu, Z.~Zhu, and S.-C.
  Cheung, ``Do the dependency conflicts in my project matter?,'' in {\em
  Proceedings of the 2018 26th ACM Joint Meeting on European Software
  Engineering Conference and Symposium on the Foundations of Software
  Engineering}, pp.~319--330, 2018.

\bibitem{McIntoshAH12}
S.~McIntosh, B.~Adams, and A.~E. Hassan, ``The evolution of java build
  systems,'' {\em Empir. Softw. Eng.}, vol.~17, no.~4-5, pp.~578--608, 2012.

\bibitem{Valero2020}
C.~{Soto-Valero}, N.~{Harrand}, M.~{Monperrus}, and B.~{Baudry}, ``{A
  Comprehensive Study of Bloated Dependencies in the Maven Ecosystem},'' {\em
  arXiv e-prints}, p.~arXiv:2001.07808, Jan. 2020.

\bibitem{styler}
B.~Loriot, F.~Madeiral, and M.~Monperrus, ``Styler: Learning formatting
  conventions to repair checkstyle errors,'' {\em arXiv preprint
  arXiv:1904.01754}, 2019.

\bibitem{Eilertsen2018}
A.~M. Eilertsen and A.~H. Bagge, ``Exploring api: Client co-evolution,'' in
  {\em Proceedings of the 2nd International Workshop on API Usage and
  Evolution}, pp.~10--13, 2018.

\bibitem{Chen2020}
L.~Chen, F.~Hassan, X.~Wang, and L.~Zhang, ``Taming behavioral backward
  incompatibilities via cross-project testing and analysis,'' in {\em IEEE/ACM
  International Conference on Software Engineering}, 2020.

\bibitem{wang2017behavioral}
Q.~Wang, Y.~Brun, and A.~Orso, ``Behavioral execution comparison: Are tests
  representative of field behavior?,'' in {\em 2017 IEEE International
  Conference on Software Testing, Verification and Validation (ICST)},
  pp.~321--332, IEEE, 2017.

\bibitem{luo2014empirical}
Q.~Luo, F.~Hariri, L.~Eloussi, and D.~Marinov, ``An empirical analysis of flaky
  tests,'' in {\em Proceedings of the 22nd ACM SIGSOFT International Symposium
  on Foundations of Software Engineering}, (New York, NY, USA), pp.~643--653,
  ACM, 2014.

\bibitem{eck2019understanding}
M.~Eck, F.~Palomba, M.~Castelluccio, and A.~Bacchelli, ``Understanding flaky
  tests: The developer’s perspective,'' in {\em Proceedings of the 2019 27th
  ACM Joint Meeting on European Software Engineering Conference and Symposium
  on the Foundations of Software Engineering}, ESEC/FSE 2019, (New York, NY,
  USA), p.~830–840, Association for Computing Machinery, 2019.

\bibitem{durieux2020flaky}
T.~Durieux, C.~Le~Goues, M.~Hilton, and R.~Abreu, ``Empirical study of
  restarted and flaky builds on travis ci,'' in {\em Proceedings of the 17th
  International Conference on Mining Software Repositories}, MSR '20, (New
  York, NY, USA), p.~254–264, Association for Computing Machinery, 2020.

\bibitem{cito2017empirical}
J.~Cito, G.~Schermann, J.~E. Wittern, P.~Leitner, S.~Zumberi, and H.~C. Gall,
  ``An empirical analysis of the docker container ecosystem on github,'' in
  {\em 2017 IEEE/ACM 14th International Conference on Mining Software
  Repositories (MSR)}, pp.~323--333, IEEE, 2017.

\bibitem{mcintosh2014mining}
S.~McIntosh, B.~Adams, M.~Nagappan, and A.~E. Hassan, ``Mining co-change
  information to understand when build changes are necessary,'' in {\em 2014
  IEEE International Conference on Software Maintenance and Evolution},
  pp.~241--250, IEEE, 2014.

\bibitem{kerzazi2014automated}
N.~Kerzazi, F.~Khomh, and B.~Adams, ``Why do automated builds break? an
  empirical study,'' in {\em 2014 IEEE International Conference on Software
  Maintenance and Evolution}, pp.~41--50, IEEE, 2014.

\bibitem{sulir2016quantitative}
M.~Sul{\'\i}r and J.~Porub{\"a}n, ``A quantitative study of java software
  buildability,'' in {\em Proceedings of the 7th International Workshop on
  Evaluation and Usability of Programming Languages and Tools}, pp.~17--25,
  2016.

\bibitem{gkortzis2020software}
A.~Gkortzis, D.~Feitosa, and D.~Spinellis, ``Software reuse cuts both ways: An
  empirical analysis of its relationship with security vulnerabilities,'' {\em
  Journal of Systems and Software}, vol.~172, p.~110653, 2020.

\bibitem{neitsch2012build}
A.~Neitsch, K.~Wong, and M.~W. Godfrey, ``Build system issues in multilanguage
  software,'' in {\em 2012 28th IEEE International Conference on Software
  Maintenance (ICSM)}, pp.~140--149, IEEE, 2012.

\bibitem{Dacapo}
S.~M. Blackburn, R.~Garner, C.~Hoffmann, A.~M. Khang, K.~S. McKinley,
  R.~Bentzur, A.~Diwan, D.~Feinberg, D.~Frampton, S.~Z. Guyer, M.~Hirzel,
  A.~Hosking, M.~Jump, H.~Lee, J.~E.~B. Moss, A.~Phansalkar, D.~Stefanovi\'{c},
  T.~VanDrunen, D.~von Dincklage, and B.~Wiedermann, ``The dacapo benchmarks:
  Java benchmarking development and analysis,'' {\em SIGPLAN Not.}, vol.~41,
  p.~169–190, Oct. 2006.

\bibitem{tomassi2019bugswarm}
D.~A. Tomassi, N.~Dmeiri, Y.~Wang, A.~Bhowmick, Y.-C. Liu, P.~T. Devanbu,
  B.~Vasilescu, and C.~Rubio-Gonz{\'a}lez, ``Bugswarm: mining and continuously
  growing a dataset of reproducible failures and fixes,'' in {\em 2019 IEEE/ACM
  41st International Conference on Software Engineering (ICSE)}, pp.~339--349,
  IEEE, 2019.

\bibitem{saha2018bugs}
R.~K. Saha, Y.~Lyu, W.~Lam, H.~Yoshida, and M.~R. Prasad, ``Bugs. jar: a
  large-scale, diverse dataset of real-world java bugs,'' in {\em Proceedings
  of the 15th International Conference on Mining Software Repositories},
  pp.~10--13, 2018.

\bibitem{durieux2016introclassjava}
T.~Durieux and M.~Monperrus, ``Introclassjava: A benchmark of 297 small and
  buggy java programs,'' 2016.

\bibitem{dyer2013boa}
R.~Dyer, H.~A. Nguyen, H.~Rajan, and T.~N. Nguyen, ``Boa: A language and
  infrastructure for analyzing ultra-large-scale software repositories,'' in
  {\em 2013 35th International Conference on Software Engineering (ICSE)},
  pp.~422--431, IEEE, 2013.

\bibitem{spinellis2020dataset}
D.~Spinellis, Z.~Kotti, and A.~Mockus, ``A dataset for github repository
  deduplication,'' in {\em Proceedings of the 17th International Conference on
  Mining Software Repositories}, pp.~523--527, 2020.

\bibitem{Leuenberger2017}
M.~Leuenberger, H.~Osman, M.~Ghafari, and O.~Nierstrasz, ``Kowalski: Collecting
  api clients in easy mode,'' in {\em 2017 IEEE International Conference on
  Software Maintenance and Evolution (ICSME)}, pp.~653--657, IEEE, 2017.

\end{thebibliography}

\end{document}